\documentclass[12pt,preprint]{aastex}

\shorttitle{The Cepheid in the binary system OGLE-LMC562.05.9009}
\shortauthors{Gieren, Pilecki, Pietrzy{\'n}ski et al.}

\begin{document}


\title{The Araucaria Project: A study of the classical Cepheid in the eclipsing binary system OGLE LMC562.05.9009 in the Large Magellanic Cloud\footnotemark[0]}

\footnotetext[0]{This research is based on observations obtained with the ESO VLT, 3.6 m and NTT telescopes for Programmes 092.D-0295(A), 091.D-0393(A), 089.D-0330(A), 
088.D-0447(A), 086.D-0103(A) and 085.D-0398(A)), and with the Magellan Clay and Warsaw telescopes at Las Campanas Observatory.}


\author{Wolfgang Gieren\altaffilmark{1,3}}
\email{wgieren@astro-udec.cl}
\author{Bogumi{\l} Pilecki\altaffilmark{2,1}}
\email{pilecki@astrouw.edu.pl}
\author{Grzegorz Pietrzy{\'n}ski\altaffilmark{2,1}}
\email{pietrzyn@astrouw.edu.pl}
\author{Dariusz Graczyk\altaffilmark{3,1}}
\email{darek@astro-udec.cl}
\author{Andrzej Udalski\altaffilmark{2}}
\email{udalski@astrouw.edu.pl}
\author{Igor Soszy{\'n}ski\altaffilmark{2}}
\email{soszynsk@astrouw.edu.pl}
\author{Ian B. Thompson\altaffilmark{4}}
\email{ian@obs.carnegiescience.edu}
\author{Pier Giorgio Prada Moroni\altaffilmark{12,13}}
\email{prada@df.unipi.it}
\author{Rados{\l}aw Smolec\altaffilmark{5}}
\email{smolec@camk.edu.pl}
\author{Piotr Konorski\altaffilmark{2}}
\email{piokon@astrouw.edu.pl}
\author{Marek G{\'o}rski\altaffilmark{3,1}}
\email{mgorski@astrouw.edu.pl}
\author{Paulina Karczmarek\altaffilmark{2}}
\email{pkarczmarkek@astrouw.edu.pl}
\author{Ksenia Suchomska\altaffilmark{2}}
\email{ksenia@astrouw.edu.pl}
\author{M{\'o}nica Taormina\altaffilmark{1}}
\email{mtaormina@astro-udec.cl}
\author{Alexandre Gallenne\altaffilmark{1}}
\email{agallenne@astro-udec.cl}
\author{Jesper Storm\altaffilmark{9}}
\email{jstorm@aip.de}
\author{Giuseppe Bono\altaffilmark{10,11}}
\email{bono@roma2.infn.it}
\author{M\'arcio Catelan\altaffilmark{7,3}}
\email{mcatelan@astro.puc.cl}
\author{Micha{\l} Szyma{\'n}ski\altaffilmark{2}}
\email{msz@astrouw.edu.pl}
\author{Szymon Koz{\l}owski \altaffilmark{2}}
\email{simkoz@astrouw.edu.pl}
\author{Pawe{\l} Pietrukowicz\altaffilmark{2}}
\email{pietruk@astrouw.edu.pl}
\author{{\L}ukasz Wyrzykowski\altaffilmark{2}}
\email{wyrzykow@astrouw.edu.pl}
\author{Rados\l{}aw Poleski\altaffilmark{2,14}}
\email{rpoleski@astrouw.edu.pl}
\author{Jan Skowron\altaffilmark{2}}
\email{jskowron@astrouw.edu.pl}
\author{Dante Minniti\altaffilmark{6,3}}
\email{dante@astrofisica.cl}
\author{K. Ulaczyk\altaffilmark{2}}
\email{kulaczyk@astrouw.edu.pl}
\author{P. Mr{\'o}z\altaffilmark{2}}
\email{pmroz@astrouw.edu.pl}
\author{M. Pawlak\altaffilmark{2}}
\email{mpawlak@astrouw.edu.pl}
\author{Nicolas Nardetto\altaffilmark{8}}
\email{Nicolas.Nardetto@oca.eu}

\altaffiltext{1}{Universidad de Concepci{\'o}n, Departamento de Astronom{\'i}a, Casilla 160-C, Concepci{\'o}n, Chile}
\altaffiltext{2}{Warsaw University Observatory, Al. Ujazdowskie 4, PL-00-478, Warszawa, Poland}
\altaffiltext{3}{Millennium Institute of Astrophysics, Chile}
\altaffiltext{4}{Carnegie Observatories, 813 Santa Barbara Street, Pasadena, CA 91101-1292, USA}
\altaffiltext{5}{Copernicus Astronomical Centre, Polish Academy of Sciences, Bartycka 18, 00-716 Warsaw, Poland}
\altaffiltext{6}{Departamento de Ciencias F{\'i}sicas, Universidad Andr{\'e}s Bello, Rep{\'u}blica 220, Santiago, Chile}
\altaffiltext{7}{Instituto de Astrof{\'i}sica, Facultad de F{\'i}sica, Pontificia Universidad Cat{\'o}lica de Chile, Av. Vicu{\~n}a MacKenna 4860, Santiago, Chile}
\altaffiltext{8}{Laboratoire Lagrange, UMR7293, Universit{\'e} de Nice Sophia-Antipolis, CNRS, Observatoire de la Cote d'Azur, Nice, France}
\altaffiltext{9}{Leibniz-Institut fur Astrophysik (AIP), An der Sternwarte 16, D-14482 Potsdam, Germany}
\altaffiltext{10}{Dipartimento di Fisica, Universita di Roma, Tor Vergata, Via della Ricerca Scientifica 1, I-00133 Roma, Italy}
\altaffiltext{11}{INAF, Rome Astronomical Observatory, via Frascati 33, I-00040 Monte Porzio Catone, Italy}
\altaffiltext{12}{Dipartimento di Fisica, Universit{\'a} di Pisa, Largo B. Pontecorvo 3, 56127 Pisa, Italy}
\altaffiltext{13}{INFN-Pisa, Largo B. Pontecorvo 3, 56127 Pisa, Italy}
\altaffiltext{14}{Department of Astronomy, Ohio State University, 140 W. 18th Avenue, Columbus, OH 43210, USA}

\begin{abstract}
We present a detailed study of the classical Cepheid in the double-lined, highly eccentric eclipsing binary system OGLE-LMC562.05.9009. 
The Cepheid is a fundamental mode pulsator with a period of 2.988 days. The orbital period of the system is 1550 days.
Using spectroscopic data from three 4-8-m telescopes and photometry spanning 22 years, we were able to derive the dynamical masses and radii
of both stars with exquisite accuracy.
Both stars in the system are very similar in mass, radius and color, but the companion is a stable, non-pulsating star. The Cepheid is slightly more massive 
and bigger ($M_{1}=3.70 \pm 0.03 M_{\odot}$, $R_{1}=28.6 \pm 0.2 R_{\odot}$) than its companion ($M_{2}=3.60 \pm 0.03 M_{\odot}$, $R_{2}=26.6 \pm 0.2 R_{\odot}$). 
Within the observational uncertainties both stars have the same effective temperature of $6030 \pm 150 K$. Evolutionary tracks place both stars
inside the classical Cepheid instability strip, but it is likely that future improved temperature estimates will move the stable giant companion
just beyond the red edge of the instability strip. Within current observational and theoretical uncertainties, both stars fit on a 205 Myr isochrone  
arguing for their common age. 

From our model, we determine a value of the projection factor of $p = 1.37 \pm 0.07$ for the Cepheid in the OGLE-LMC562.05.9009 system. This is the second Cepheid for which we could
measure its p-factor with high precision directly from the analysis of an eclipsing binary system, which 
represents an important contribution towards a better calibration of Baade-Wesselink methods of distance determination for Cepheids.

\end{abstract}


\section{Introduction}  
\label{sect:intro}

Classical Cepheid variables are of great importance in astrophysics. They obey the famous period-luminosity relation (Leavitt 1908),
now also named the "Leavitt Law", which has made them excellent standard candles to calibrate the first rungs of the extragalactic distance scale in the
local Universe (e.g. Gieren et al. 2005a, Freedman \& Madore 2010, Riess et al. 2011, Kodric et al. 2015). Our currently most accurate approach to determine
the Hubble constant uses a distance scale building on classical Cepheids in tandem with Ia-type Supernovae (e.g. Riess et al. 2011). 
Cepheids are also excellent tools to check on the validity, and improve stellar pulsation and stellar evolution theories (Caputo et al. 2005).
One of the serious and long-standing problems of these theories was an inconsistency, at a level of $\approx 20-30\%$, between the masses
predicted by the evolutionary and pulsational routes
(Stobie 1969; Cox 1980; Keller 2008; Neilson et al. 2011 and references therein). The obvious way to solve this "mass discrepancy problem" was to find 
Cepheid variables in double-lined eclipsing binary systems which would allow to accurately determine their dynamical masses. However, it took more than 40 years until
such a system (OGLE-LMC-CEP-0227) was finally found by Soszy{\'n}ski et al. (2008) and subsequently studied
by our group, yielding the dynamical mass of the Cepheid 
with an exquisite accuracy of 1\% (Pietrzynski et al. 2010; Pilecki et al. 2013). This dynamical mass determination has already led to
improvements in stellar evolution (Cassisi and Salaris 2011, Neilson et al. 2011, Prada Moroni et al. 2012, Neilson and Langer 2012), and stellar 
pulsation theories (Marconi et al. 2013).

The detection of a Cepheid in a double-lined eclipsing system not only allows to determine its mass with excellent accuracy, but also allows 
to measure highly accurate values of other physical parameters which are impossible to determine, with a similar accuracy, for a single Cepheid,
or a Cepheid in a non-eclipsing binary system. Our previous work has shown that radii accurate to 1-3\% can be obtained, depending on the
configuration of the system components at the primary and secondary eclipses (Pilecki et al. 2013, 2015). Such an accurate 
radius determination poses a strong constraint on the pulsation mode of the Cepheid. The orbital and photometric solutions also allow to determine the p-factor
of the Cepheid which is needed in Baade-Wesselink (BW)-type distance determinations of Cepheids to convert their measured radial velocities
to the pulsational velocities of the Cepheid surfaces, and currently constitutes the largest source
of systematic uncertainty in any type of BW analyses (Storm et al. 2004; Gieren et al. 2005b; Fouqu{\'e} et al. 2007; Storm et al. 2011). A direct
and accurate measurement of the p-factors for a number of Cepheids spanning a range of pulsation periods will be of enormous value in the effort
to achieve distance determinations accurate to 1-3\% for single Cepheids with the BW method. Apart from the p-factor, the full analysis
of a Cepheid-containing eclipsing binary, including the analysis of high-quality NIR data, is also able to provide a precise estimation of the limb darkening of the Cepheid 
(see Pilecki et al. 2013;
hereafter P13), which cannot be determined empirically in any other way. 

The object of this study, OGLE LMC562.05.9009, was discovered as an eclipsing binary with a Cepheid component from OGLE IV data by Soszy{\'n}ski et al. (2012) 
in the OGLE South Ecliptic Pole LMC fields prepared with the aim of providing tests for the Gaia satellite mission. No orbital period
for the system could be derived from these data however. It is not in the list of Cepheids discovered
by the MACHO project (Alcock et al. 2002), but is contained in the list of Cepheids published by the EROS-2 group (Kim et al. 2014). Its name in
the EROS-2 database is lm0240n14595, but there was no information about eclipses. Given the discovery of Soszy{\'n}ski et al. (2012) of eclipses,
we initiated extensive
high-resolution spectroscopy and
follow-up photometry of the LMC562.05.9009 system (see section 2 of this paper), which led to the spectroscopic confirmation of its genuine binary
nature, and eventually allowed an accurate determination of its orbital period. With the photometric data, and an extensive catalog of radial velocity
observations of the system, we were able to precisely disentangle the pulsational and orbital radial velocity variations, and provide full and accurate
orbital and photometric solutions of the system, following the methodology which was used by P13 in the analysis
of the OGLE-LMC-CEP-0227 system. This results in the determination of accurate physical parameters for both the Cepheid and its non-pulsating binary
companion. Future near-infrared photometric coverage of the next eclipses, which will occur in about four years from now, will improve on the characterization 
of the physical parameters of the two stars. 

Our paper is organized as follows. In section 2, we describe the data underlying this study. In secion 3, we will present the data analysis methods, and the results emerging
from our analysis of the observational data. In section 4, conclusions and an outlook on future work will be presented.

\section{Data}
\label{sect:data}

In order to reliably and accurately separate the orbital motion from the radial velocity variations due to the pulsation of the Cepheid component
in the system, a large number of precise radial velocity measurements, providing good phase coverage of both the pulsational, and orbital radial
velocity curves, was necessary. This task was not made easier by the close-to-integer value of 2.988 days of the pulsation period of the Cepheid. We obtained
high-resolution echelle spectra using the UVES spectrograph at the ESO-VLT on Paranal (49 epochs), the MIKE spectrograph at the 6.5-m Magellan Clay
telescope at Las Campanas Observatory (22 epochs), and the HARPS spectrograph at the 3.6-m telescope at ESO-La Silla (10 epochs). The UVES data were reduced
using a standard ESO pipeline and software obtained from the ESO webpage (http://www.eso.org/sci/software.html) (Freudling et al. 2013). The MIKE data were reduced
with the pipeline software written by Dan Kelson, following the approach outlined in Kelson (2003). The HARPS data were reduced on-site by the Online Reduction System.

Radial velocities were measured using the Broadening Function method (Rucinski 1992, 1999) implemented in the RaveSpan software (Pilecki et al. 2012). Measurements 
were made in the wavelength interval 4125 to 6800 $\mathring{A}$ which contains numerous metallic lines. Synthetic spectra taken from the
library of Coelho et al. (2005) were used as templates. The typical formal errors of the derived velocities are $\sim 370$ m/s. The individual radial velocity
measurements for both components of the OGLE LMC562.05.9009 system are available online at: 

\centerline{ http://araucaria.astrouw.edu.pl/p/cep9009 }
 
In some cases where line profiles of both companions were blended, only the velocity of the Cepheid (number 1 in the table) was measured. By fitting the systemic
radial velocities with the datasets from the different instruments, we found offsets of +250 m/s for MIKE,
and -210 m/s for HARPS, respectively with respect to the UVES radial velocity system. These small offsets have been taken into account. Even
if we would not have corrected for these very small velocity shifts between the different instruments, the orbital solution and the physical parameters of the 
component stars derived in the following sections would not have changed in any significant way.

A total of 588 photometric measurements in the $I$-band and 143 in the $V$-band were collected with the Warsaw telescope by the OGLE project (Udalski et al. 2015), 
and during observing time granted to the Araucaria project by the Chilean National Time Allocation Committee (CNTAC). The images were reduced with
the OGLE standard photometric pipeline based on difference image analysis, DIA (Udalski et al. 2015), and instrumental magnitudes were calibrated onto
the standard system using Landolt standards. The typical accuracy of the measurements was at the
5 mmag level. We have also used instrumental $V$-band and $R$-band data 
from the MACHO project (ID: 71.11933.15) downloaded from the webpage http://macho.anu.edu.au and converted to the Johnson-Cousins system using equations from 
Faccioli et al.(2007). We augmented our data with $R_{EROS}$ (equivalent to Johnson-Cousins I) data from the EROS project 
(Kim et al. 2014)\footnote{http://stardb.yonsei.ac.kr}. While the very precise  OGLE data are crucial for our analysis,
the inclusion of the MACHO and EROS data in our study was important for an accurate determination of the orbital period of the OGLE LMC562.05.9009 system,
and to improve the coverage of the secondary eclipse in the light curve. They also helped to considerably improve the pulsational V-band
light curve of the Cepheid. The
MACHO $V$-band and EROS data were shifted in flux and magnitude to fit the OGLE data by the minimization of the difference between the out-of-eclipse light curves. 
This way the light curves were forced to have the same average magnitude. The MACHO $R$-band light curve was left unmodified. The flux shift 
was later modeled by adding a third light in this band which simulates a flux shift, which may appear due to any calibration error.

The individual photometric data are given on the same webpage as the radial velocity data (see above), and
are shown in Figure 1. In Figure 2, we show the pulsational light curves of the Cepheid in the I, R and V bands, as obtained from the out-of-eclipse photometric data
folded with the pulsation period of the star. These data resemble light curves of very low scatter with an asymmetrical shape typical for a Cepheid pulsating in the 
fundamental mode. The Fourier decomposition parameters of the I-band light curve of the Cepheid, shown in Figure 3, clearly confirm that the star is indeed 
a fundamental mode pulsator.

The pulsation period of the Cepheid is very accurately determined from the current data. From the V band data, we obtain a period of 2.9878463 (09) days,
while the I band data yield a period of 2.9878466 (16) days, leading to the uncertainty of the pulsation period quoted in Table 3 which is 
consistent with the absolute value of rate of period change
being $<$0.1 s/yr. The O-C diagrams for both the I and V band data do not display any secular systematic change, confirming that the total uncertainty
on the period as given in Table 3 is correctly estimated.

Figure 4 shows the orbital light curve of the system for the OGLE IV I-band data, folded on the orbital period of 1550.4 days, and with the 
pulsational variations of the Cepheid removed. It is seen that the orbit of the LMC562.05.9009 system is highly eccentric, and that both the primary and secondary eclipses 
are covered by the data.

In order to determine the effective temperatures of the component stars, we augmented our dataset with 12 epochs of J and K photometry which were obtained outside
the eclipses. These data were taken with the SOFI near-infrared camera attached to the ESO NTT 3.5 m telescope on La Silla. The reduction and calibration of the data
to the UKIRT system (Hawarden et al. 2001) was done following the procedure described in detail in Pietrzy{\'n}ski et al. (2006).
The accuracy of the zero points in both bands is 0.015 mag, and instrumental errors are not larger than 0.01 mag.

\section{Analysis and Results}
\label{sect:results}

>From the analysis of the radial velocity curve of a binary star one can obtain the orbital parameters of the system. In the case of the studied system the procedure 
is complicated by the pulsational variability of the Cepheid superimposed on the orbital motion. Using the RaveSpan software we have fitted a model of Keplerian orbit 
(i.e. proximity effects were ignored, which is justified by the large distance of the components even at closest approach) with an additional Fourier series 
representing the pulsational radial velocity curve of the Cepheid.

We simultaneously fitted the reference time $T_0$, the eccentricity $e$, the argument of periastron $\omega$, the velocity semi-amplitudes $K_1$ and $K_2$, 
the systemic velocity $\gamma$, and Nth-order Fourier series. In the beginning systemic velocities of both components were fitted, but without any improvement 
in the fit and with the values equal within the errors. Eventually only one velocity was kept.

The period $P$ was initially held fixed at the estimated value of 1550.4 days. The fitting was later repeated with a fixed value of $P=1550.354$ d and $T_0$ calculated 
from the photometric epoch of a primary minimum $T_{I}$ as a function of eccentricity and argument of periastron:
$$T_0 = f(e, \omega; P=1550.354~d, T_{I}=3959.23~d)$$
to ensure the consistency of the model.

The error of the eccentricity turned out to be 10 times higher and the error for the argument of periastron 6 times lower (see solution 3 in Table~\ref{tab:spec}) 
than the ones obtained from the photometry. For this reason we have tried to solve the system with $e$ and $\omega$ fixed (solution 1), or only $e$ fixed (solution 2). 
Eventually we decided to adopt solution 2 as the final one because of the low error for $e$ from the photometry and the low error of $\omega$ from the orbital solution. 

In this way we have obtained the coefficients describing the pulsational radial velocity curve and the parameters describing the orbital motion separately. 
The orbital radial velocity curve along with the best fitting model is shown in Fig.~\ref{fig:rvorb}.
To obtain the pulsational radial velocity curve of the Cepheid we then subtracted the orbital motion from the measured velocities. The resulting radial velocity curve is shown 
in Fig.~\ref{fig:rvpuls} together with the radius variation curve calculated with the $p$-factor 1.37 obtained from the fit. The orbital solutions are presented 
in Table~\ref{tab:spec}.

The photometric data were analyzed using a version of the JKTEBOP code (Popper \& Etzel 1981, Southworth et al. 2004, 2007) modified to allow the inclusion of 
pulsation variability. We have previously used this package in the analysis of the OGLE-LMC-CEP-0227 system (P13), and we refer the reader 
to this work for more details.

We varied the following parameters in deriving the final model: the fractional radius of the pulsating component at phase 0.0 (pulsational), $r_1$;
the fractional radius of the second component, $r_2$; the orbital inclination $i$; the orbital period, $P_{orb}$; the epoch of the primary minimum, $T_{I}$; 
the component surface brightness ratios in all three photometric bands at phase 0.0 (pulsational), $j_{21}$; and the third light in the $R_C$-band $l_{3} (R_C)$. 
The radius change 
of the Cepheid was calculated from the pulsational radial velocity curve using the $p$-factor value of 1.37 and the change of the surface brightness ratios from 
the instantaneous radii and out-of-eclipse pulsational light curves (for details, see P13). The third light in the $R_C$-band was introduced because we were unable 
to transform it directly to the OGLE photometric system.

The search for the best model (lowest $\chi^2$ value) was made using the Markov chain Monte Carlo (MCMC) approach (Press et al. 2007) as described in P13. 
The best fit photometric parameters are presented in Table~\ref{tab:photpar}. We present two photometric solutions in this Table. In the first one, the argument
of periastron $\omega$ is taken from the orbital solution, in the second one it is fitted. We consider Solution 1 as the final one, being consistent with
the above discussion of the $\omega$ errors. In this way we take the best advantage from the photometric and orbital radial velocity data.
Using these parameters we generated a model for each light curve. 
In Fig.~\ref{fig:ieclmodel} we show a close-up of selected eclipses for each passband. The magnitude range is the same for all plots to facilitate the comparison.

Most of the parameters fitted in our approach are independent and do not exhibit any significant correlation. The only significant correlation is between 
the orbital plane inclination $i$ and the sum of the radii $r_{1}+r_{2}$ as shown in Fig.~\ref{fig:corr_i}.

\subsection{Eclipses}
In order to better understand the configuration of the system using the derived parameters we calculated the distances between the stars at important phases. 
At the phase of the primary eclipse the distance between the components is about $650 R_{\odot}$, while at the phase of the secondary eclipse it is 
about $725 R_{\odot}$. Both eclipses occur when the stars are relatively close to each other. The minimum and maximum separations during the orbital cycle 
are 425 and 1760 $R_{\odot}$, respectively.
At the primary eclipse the projected distance between the centers of the stars is $22.9 R_{\odot}$, and at the secondary eclipse the projected distance is $25.6 R_{\odot}$, 
while the sum of the radii changes between 53.4 and 56.4 $R_{\odot}$ depending on the instantaneous radius of the Cepheid. The configuration at both phases 
is illustrated in Fig.~\ref{fig:config}.

\subsection{Radius and projection factor}
To test the results of our analysis, we have calculated the expected radius of the Cepheid from period-radius (PR) relations for classical Cepheids in the literature. 
The relation 
of Gieren et al. (1998) for fundamental mode pulsators yields an expected mean radius value of $27.0 \pm 1.2 R_{\odot}$ for the pulsation period of the Cepheid in our system
which agrees with our determination ($28.6 \pm 0.2 R_{\odot}$) within the combined 1 $\sigma$ errors. The PR relations of Sachkov (2002) for fundamental mode
and first overtone Cepheids predict radii of $27.4 \pm 0.9 R_{\odot}$ and $35.6 \pm 5.4 R_{\odot}$, respectively, for a pulsation period of 2.988 days. The first value 
matches our derived radius value for the OGLE LMC562.05.9009 Cepheid much better, and is in agreement with the radius prediction from the Gieren et al. (1998) PR relation. 
We conclude that the radius value of the Cepheid clearly supports fundamental mode pulsation, in agreement with the conclusion reached
from the Fourier decomposition parameters of the I-band light curve.
The radius value together with the other physical parameters of the Cepheid given in Table 3, particularly its mass, 
also leave no doubt that the pulsating star in the system is a classical (and not a Type-II) Cepheid.

Our models constrain the projection factor of a Cepheid in an eclipsing binary system in the way which has been discussed in detail in P13. Briefly,
the shape of a Cepheid light curve in a given photometric band is determined by the change of its surface temperature and its radius. The radius change is
particularly important if the Cepheid resides in an eclipsing binary system. The beginning and end of an eclipse may be shifted in time according to the
instantaneous radius of the Cepheid, and the visible area of the eclipsed stellar disk depends on the phase of the pulsating component. In our approach
the Cepheid variability is a part of the model, so we can trace the influence of the related parameters on the light curve. As a base we use the raw (unscaled) 
absolute radius change obtained from the pulsational radial velocity curve. Then we scale its amplitude with the projection factor (the $p$-factor scales
linearly with the amplitude of the radius variation curve). A conversion from the absolute radii to the relative radii (used in the light curve analysis)
is done by using the orbital solution. A comparison of the resulting model light curves with the data then directly constrains the $p$-factor value. From
our best model we obtain a radius variation amplitude of 3.04 R$_\odot$ for the Cepheid, which corresponds to $p=1.37$ (see Figs. 6 and 9).
 
Our current determination of the projection factor of the Cepheid in the OGLE LMC562.05.9009 system is the second reliable measurement of this important quantity
for a Cepheid in a binary, after the first determination made by P13 for OGLE-LMC-CEP-0227. The value of $p = 1.37 \pm 0.07$ is smaller than the predicted p-factor value
from the most recent calibration of the p-factor relation of Storm et al. (2011) which yields $p = 1.46 \pm 0.04$ for the pulsation period of the Cepheid. However, there is possible 
agreement within the combined uncertainties of the two values. This is contrary to the finding for CEP-0227 which has a pulsation period of 3.80 days 
and $p = 1.21 \pm 0.04$ from
our analysis in P13, whereas its expected p-factor value from the Storm et al. calibration is $p = 1.44 \pm 0.04$, with both values clearly discrepant within their respective
uncertainties. The large difference of the p-factor values for the two binary Cepheids for which we could determine
this number so far with our method is also noteworthy (the difference is 0.16, whereas the p-factor relation of Storm et al. predicts a difference of only 0.02 for a change of the period 
from 2.988 to 3.80 days. Other p-factor relations, such as the theoretical relations of Neilson et al. (2012), predict an even smaller change of $p$ between the two
period values). Our finding hints at the possibility that the p-factor - period relation may have an intrinsic dispersion, particularly in the short pulsation period range,
where the discrepancy of the p-factor values predicted by different calibrations of the relation in the literature is largest (see discussion in Storm et al. 2011
and Gieren et al. 2013). 

\subsection{Extinction and temperature}

The extinction in the direction to the target was calculated in a similar way as described in Pilecki et al.(2015). We utilized the observed 
(not extinction-corrected) period-magnitude relations for fundamental mode Cepheids in the LMC (Soszy{\'n}ski et al. 2008) in the optical V and I bands.
By comparing the observed mean magnitudes of the Cepheid with the expected magnitudes for its period,
 we determined the differential color excess (with respect to the LMC mean value) as $\Delta E(B\!-\!V)=-0.016$ mag, and a total color excess 
 of $E(B\!-\!V)=0.106$ mag using the mean extinction for the LMC given by Imara \& Blitz (2007) - see Table~\ref{tab:magnitud}. This color excess corresponds 
 to a total extinction in the K-band of $A_K=0.036$ mag.

The mean (over the pulsation cycle of the Cepheid) observed IR magnitudes of the OGLE LMC562.05.9009 system are $J=14.232\pm 0.018$ and $K=13.873\pm 0.018$ mag. 
They were transformed 
onto the 2MASS system using the equations of Carpenter (2001). We calculated an expected exctinction-free K-band magnitude of the Cepheid using relations 4 and 13 from 
Ripepi et al.(2012). The observed and de-reddened magnitudes of both components in the $V$, $I_C$ and $K$ bands are given in Table~\ref{tab:magnitud}. 
The effective temperatures of the two stars were then calculated from their intrinsic colors, using the calibrations by Worthey \& Lee (2011). 
The extinction-corrected magnitudes and colors of the primary and secondary components and their temperatures are given in Table 3.

It is very interesting to note that within the uncertainties both components have the same effective temperatures, luminosities and surface gravities. 
However, according to the very precise OGLE-IV photometry the secondary does not show any pulsations with amplitude larger than 0.01 mag.
This is a striking result 
because, assuming the same chemical composition for both components in the system, we would expect both stars to be located 
within the instability strip (see discussion next section). The fact that the secondary is non-pulsating and thus outside the instability strip could imply
that the two components of OGLE LMC562.05.9009 have significantly different abundances, which would make this system 
unique among known binary stars. 

A different, and probably more likely explanation is that the secondary is just a little cooler that the Cepheid ($\sim 70$ K), as suggested by 
our photometric Solution 2 in Table 2. In that case the Cepheid would reside almost exactly on the red boundary of the instability strip, with the secondary 
located just beyond the red edge. If this scenario is the correct one, the present work would provide the best known observational constraint on the exact position 
of the instability strip red edge. 

\subsection{Evolutionary status and age of the Cepheid and its companion}

We computed the evolutionary tracks of the two component stars of the OGLE LMC562.05.9009 system by means of the Pisa release of the FRANEC code
(Degl'Innocenti et al. 2008; Tognelli et al. 2011) adopting the same input physics and prescriptions described in detail in Dell'Omodarme et al. (2011).
An important exception is the neglecting of microscopic diffusion of helium and metals, because of their negligible impact on the evolution of
intermediate-mass stars, as we did in our previous paper on OGLE-LMC-CEP-0227 (Prada Moroni et al. 2012). During the central hydrogen burning phase,
we took into account an overshooting of $l_{ov}$=$\beta_{ov} H_p$ - where $H_p$ is the pressure height-scale and $ \beta_{ov}=0.25 $ - beyond the Schwarzschild 
classical border of the convective core. We computed the evolutionary tracks and isochrones adopting a value of the mixing-length parameter - which
parametrizes the efficiency of the super-adiabatic convection - $\alpha$ = 1.74. This value results from a solar calibration with our own
Standard Solar Model computed with the same version of the FRANEC code used to compute the evolutionary tracks in this work. For a quantitative evaluation
of some of the main sources of uncertainty affecting the theoretical evolutionary models of He-burning stars of intermediate mass we refer to
Valle et al. (2009).

The initial metal and helium abundances adopted for the calculations are Z=0.005 and Y=0.258, respectively.

In Figure 10, the locations of the two stars on the luminosity-effective temperature diagram from the parameters derived in this study
(see Table 3) are shown. Also plotted are the boundaries of the classical fundamental mode Cepheid instability strip, for metallicities of Z=0.004
and Z=0.008, taken from Bono et al. (2005). It is seen that for both metallicities, not only the Cepheid, but also the stable companion star
are located inside the instability strip. A likely explanation is that the current uncertainty on the effective temperature of the companion star 
is somewhat underestimated and that a future, more accurate determination of the temperature will move the non-pulsating star in OGLE LMC562.05.9009
slightly beyond the Cepheid instability strip; but there is also the possibility of significant different metallicities of the two stars.

Also shown in Fig. 10 are the evolutionary tracks computed for the masses of the two stars, using the prescripts detailed above. A isochrone for an age of
205 Myr fits the position of both stars on the diagram reasonably well within the observational uncertainties on their luminosities and temperatures.
The age of the Cepheid expected from the theoretical period-age relation for fundamental mode classical Cepheids of Bono et al. (2005, their Table 4) 
for a metallicity of Z=0.004 (very slightly smaller than our assumed metallicity of Z=0.005 for the calculation of the isochrone) is $130 \pm 35 Myr$. Our
current age determination for the classical Cepheid in the binary system is about $2\sigma$ larger than its age as predicted from the Bono et al.
period-age relation, but given the uncertainties involved the two values are marginally consistent. We will check on this more deeply once we have
new data which will allow us a more accurate determination of the temperatures and luminosities of the two stars, and of their metallicities, leading to a more
accurate age determination from the isochrone method. The current
results do however support the conclusion that both stars in the OGLE LMC562.05.9009 system are coeval, with an age larger than, but within the errors consistent with
the value predicted for the Cepheid from a theoretical period-age relation.

\section{Conclusions}
\label{sect:concl}

We have confirmed from high-resolution spectra that the eclipsing binary system OGLE LMC 562.05.9009 contains a classical Cepheid pulsating
with a period of 2.988 days in orbit with a stable secondary component. We performed the analysis
of our extensive spectroscopic and photometric datasets in the same way as described in our previous analysis of the OGLE-LMC-CEP-0227
system by P13, and have derived very accurate masses (to 0.8\%) and radii (0.7\%) for both the Cepheid and its non-pulsating companion star, which has a nearly identical
mass and radius as the Cepheid. The orbit is highly eccentric with $e=0.61$ and a very long period of 1550 days, or 4.2 years. Our solution defines the orbital
radial velocity curves of both components, disentangled from the pulsational velocity variations of the Cepheid, extremely well, as well as the pulsational radial
velocity curve of the Cepheid. Our analysis yields the second precise determination of the p-factor of a Cepheid in a binary so far in the literature,
and was used to determine the radius variation of the Cepheid over its pulsation cycle. Our model reproduces the observed light curves extremely well, particularly the
primary eclipse when the companion star transits in front of the Cepheid. We calculated evolutionary tracks for the two component stars in the system
and find that a isochrone for an age of 205 Myr fits the observed positions of both stars in the luminosity-effective temperature plane, arguing for the same age
of the Cepheid and its red giant companion.

The p-factor value for the Cepheid is marginally consistent with the prediction of the p-factor relation of Storm et al. (2011), as opposed to the p-factor we derived 
for OGLE-LMC-CEP-0227 in P13, which is in significant disagreement with the prediction of the Storm et al. relation. Currently the situation regarding the correct p-factor
values to use in Baade-Wesselink-type Cepheid distance determinations is still very confusing. The measurements from the two binary Cepheids in this paper
and in P13 seem to support the idea that the p-factor for classical Cepheids is not only period-dependent, but might also possess an intrinsic dispersion,
at least for short pulsation periods in the range of a few days. Clearly more work is needed to clarify this question, and one of the very few observational approaches
which promise to solve the issue is the analysis of more Cepheids in eclipsing binaries whose characteristics allow the determination of their p-factors.
The most
important parameter in this context is the radius variation amplitude of the Cepheid; the larger the amplitude, the stronger the effect of the radius variation on 
the binary light curve,
and the smaller the uncertainty on the p-factor derived from our model. This was the reason why we could not measure the p-factor for the first overtone Cepheid
in the eclipsing system OGLE-LMC-CEP-2532 whose radius variation is too small to cause a significant effect on the binary light curve, given
the quality of the photometric data (Pilecki et al. 2015).
Since fundamental mode Cepheids tend to have larger radius variations, precise measurements of Cepheid projection factors with our binary method will mostly be restricted
to eclipsing systems containing fundamental mode Cepheids.

In order to analyze the OGLE LMC562.05.9009 system, and in particular its Cepheid more fully, we plan to observe more eclipses (both primary and secondary) in the
future, including coverage in near-infrared bands. A high quality out-of-eclipse the pulsational K-band
light curve of the Cepheid in tandem with the V-band light and pulsational radial velocity curves as determined in this paper will allow us to calculate
the distance to the Cepheid with the BW-type Infrared Surface Brightness Technique (Fouqu{\'e} \& Gieren 1997; Storm et al. 2011) and compare it to the distance
of its companion star determined from the binary analysis and a surface brightness-color relation, as described in Pietrzynski et al. (2009, 2013). Such a comparison
will put further constraints on the p-factor relation valid for classical Cepheid variables.

Our work has now revealed and analyzed the fifth eclipsing binary system containing a classical Cepheid in orbit with a stable giant star. Previous binary Cepheids
analyzed by our group are OGLE-LMC-CEP-0227 (Pietrzynski et al. 2010, P13), OGLE-LMC-CEP-1812 (Pietrzynski et al. 2011), OGLE-LMC-CEP-1718 (Gieren et al. 2014), and
OGLE-LMC-CEP-2532 (Pilecki et al. 2015). The most exotic system is OGLE-LMC-CEP-1718 which contains two classical Cepheids in a 413-day orbit. Its analysis in
Gieren et al. (2014) has been very challenging due to the multiple superimposed variations in the light- and radial velocity curves. We hope to improve
on the analysis of that exciting system in the near future with additional data and possible improvements in our analysis code. For all systems but one,
OGLE-LMC-CEP-1812, the mass ratio is very close to, or consistent with unity. The exception in the case of OGLE-LMC-CEP-1812 is probably explained by the result 
reported by Neilson et al. (2015a) that the Cepheid in that system is actually the product of a stellar merger of two main sequence stars. From an observational 
point of view, there is a bias which favors the finding of systems composed of a Cepheid in orbit with a giant star of similar mass and radius which leads not only to
a higher probability to observe both eclipses, but also to observe the lines of both components in the composite spectra. For this reason, we cannot argue
that our results to-date on Cepheids in double-lined eclipsing binary systems in the LMC contradict results regarding the binary distribution of Cepheids
as obtained by Evans et al. (2015), or Neilson et al. (2015b).

The binary Cepheids in the LMC, with their dynamical masses determined to better than 2\%, will be a cornerstone for
improving our detailed understanding of Cepheid pulsation and post-main sequence stellar evolution, and in general of our understanding of Cepheid physics. With future
precise distance determinations to these systems we hope to determine from the stable binary companions, these binary Cepheids will also become excellent
absolute calibrators of the extragalactic distance scale.

\acknowledgments
We gratefully acknowledge financial support for this work from the BASAL Centro de Astrof{\'i}sica y Tecnolog{\'i}as Afines (CATA) PFB-06/2007, from
the Polish National Science Center grant MAESTRO DEC-2012/06/A/ST9/00269, and from the Polish NCN grant DEC-2011/03/B/ST9/02573.
WG, MG, DG, DM and MC also gratefully acknowledge support for this work from the Chilean Ministry of Economy, Development and Tourism's Millennium Science Initiative through grant IC120009 
awarded to the Millennium Institute of Astrophysics (MAS). AG acknowledges support from FONDECYT grant 3130361, and MC from FONDECYT grant 1141141.
The OGLE Project has received funding 
from the National Science Center, Poland, grant MAESTRO 2014/14/A/ST9/00121 to AU.


This paper utilizes public domain data obtained by the MACHO Project, jointly funded by the US Department of Energy through the University of California, Lawrence Livermore National Laboratory
 under contract No. W-7405-Eng-48, by the National Science Foundation through the Center for Particle Astrophysics of the University of California under cooperative agreement AST-8809616, 
 and by the Mount Stromlo and Siding Spring Observatory, part of the Australian National University.

We would like to thank the support staffs at the ESO Paranal and La Silla and Las Campanas Observatories for their help in obtaining the
observations.
We thank the ESO OPC and the CNTAC for generous allocation of observing time for this project.

This research has made use of NASA's Astrophysics Data System Service.

{\it Facilities:} \facility{ESO:3.6m (HARPS)}, \facility{ESO:NTT (SOFI)}, \facility{VLT:Kueyen (UVES)}, \facility{Magellan:Clay (MIKE)}, \facility{Warsaw telescope}.

\begin{figure}
\begin{center}
  \resizebox{\linewidth}{!}{\includegraphics{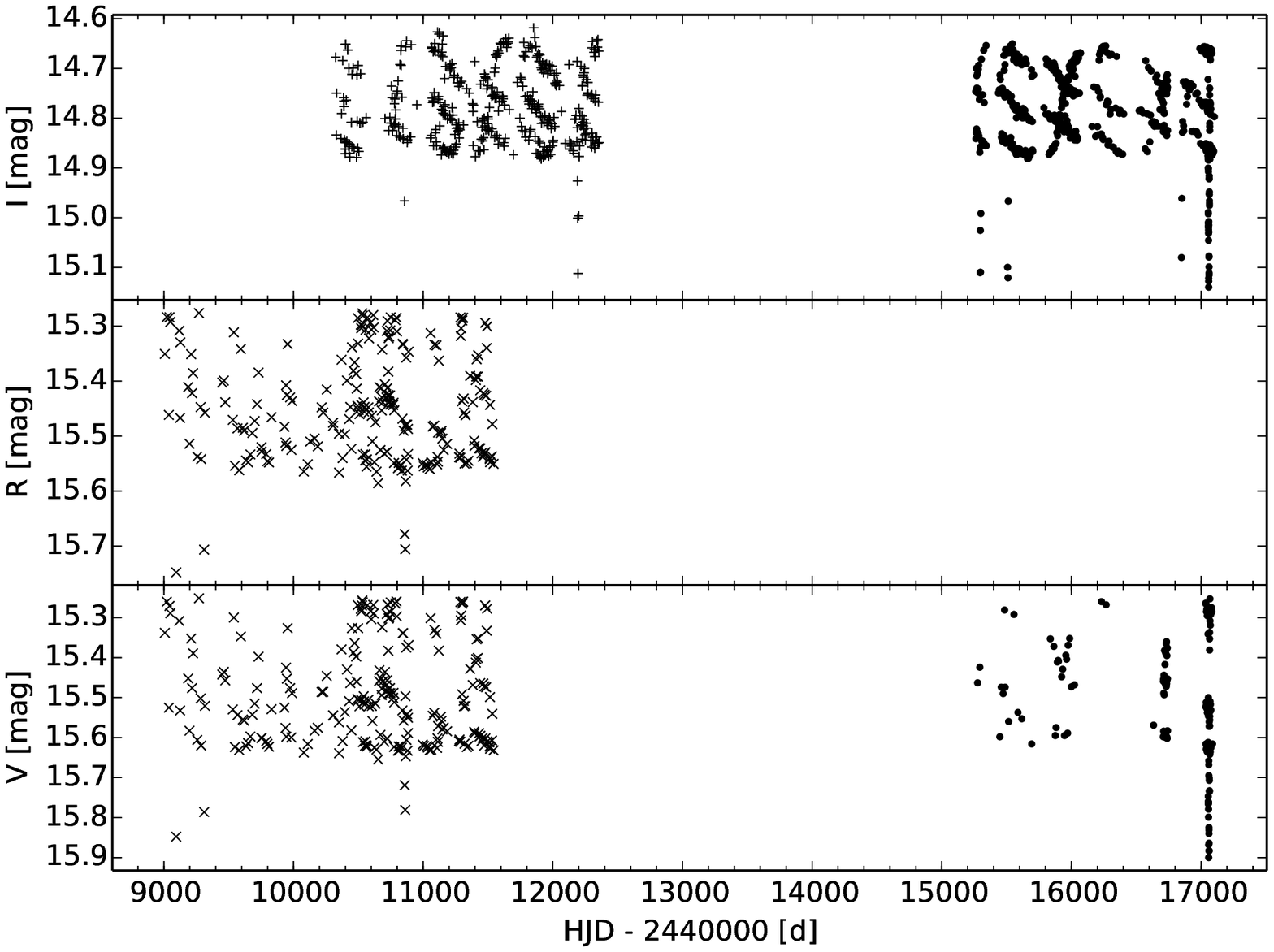}} \\
\caption{Photometric data collected for LMC562.05.9009. {\it Upper panel:} OGLE (dots) and converted EROS-R (+) $I$-band data; {\it middle panel:} MACHO $R$-band data; {\it lower panel:} OGLE (dots) and MACHO (x) $V$-band data.
The pattern seen in the time series out-of-eclipse data is due to the near-3 day pulsation period of the Cepheid in the system.}
\label{fig:obs_all}
\end{center}
\end{figure}


\begin{figure}
\begin{center}
  \resizebox{\linewidth}{!}{\includegraphics{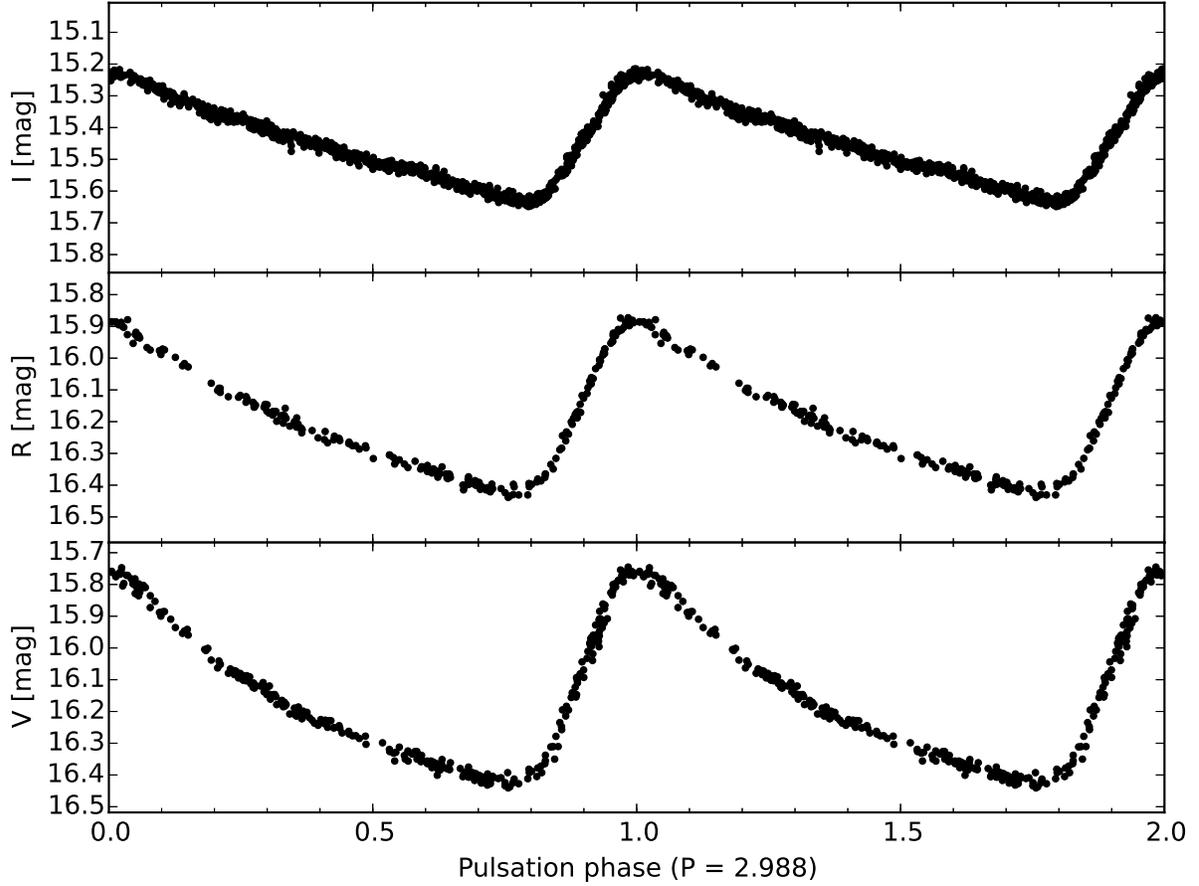}} \\
\caption{Out-of-eclipse pulsational light curves of the Cepheid in the LMC562.05.9009 system, freed from the companion light and folded 
with the ephemeris $T_{max} ($HJD$) = 2454507.90 + 2.987846\times $E. The magnitudes in this figure have not been dereddened.
The Y-axis span is always the same (0.85 mag) -- one can see that the amplitude is smaller for the redder filters.
\label{fig:lcpuls}}
\end{center}
\end{figure}

\begin{figure}
\begin{center}
  \resizebox{0.6\linewidth}{!}{\includegraphics{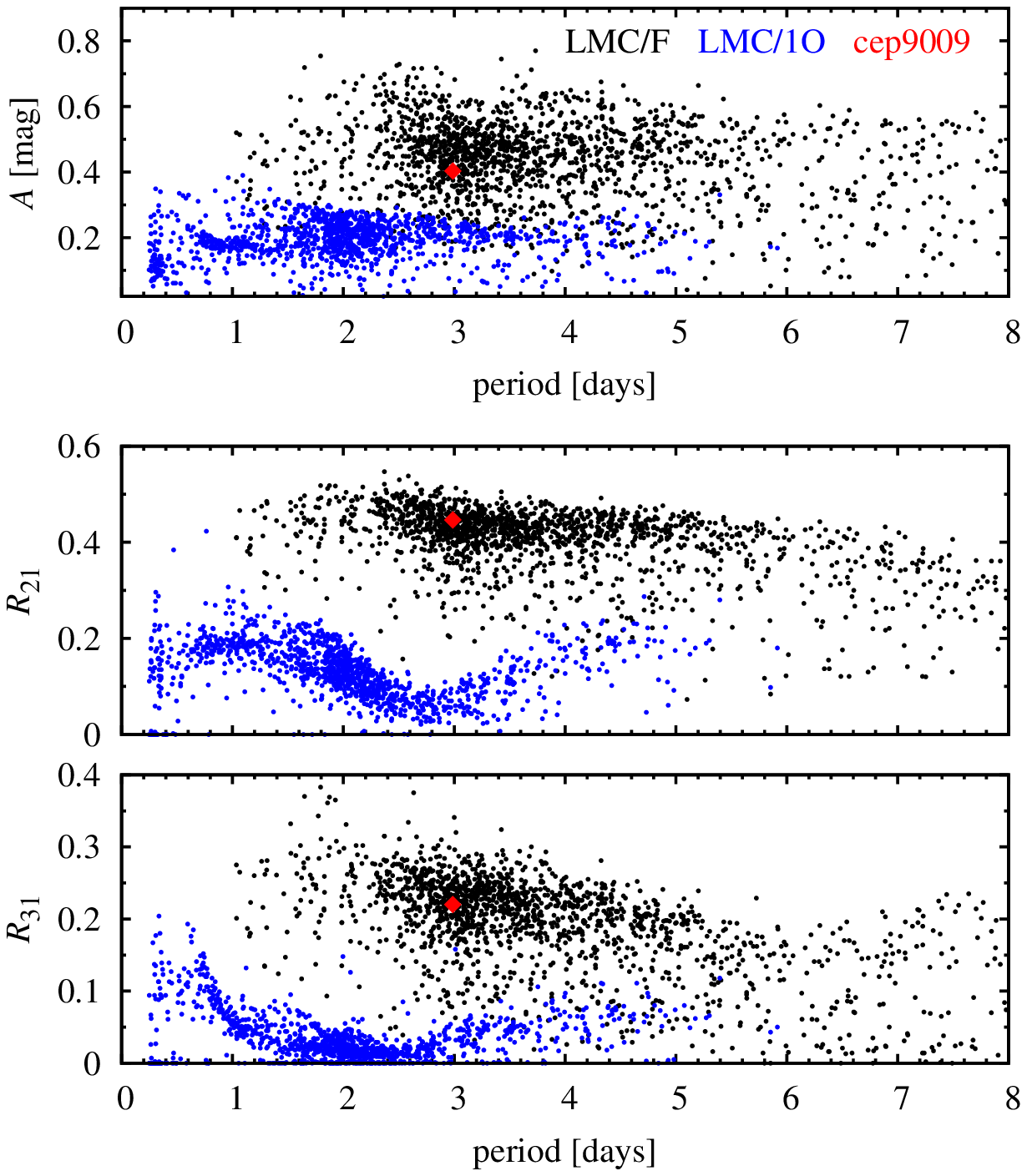}} \\
  \resizebox{0.6\linewidth}{!}{\includegraphics{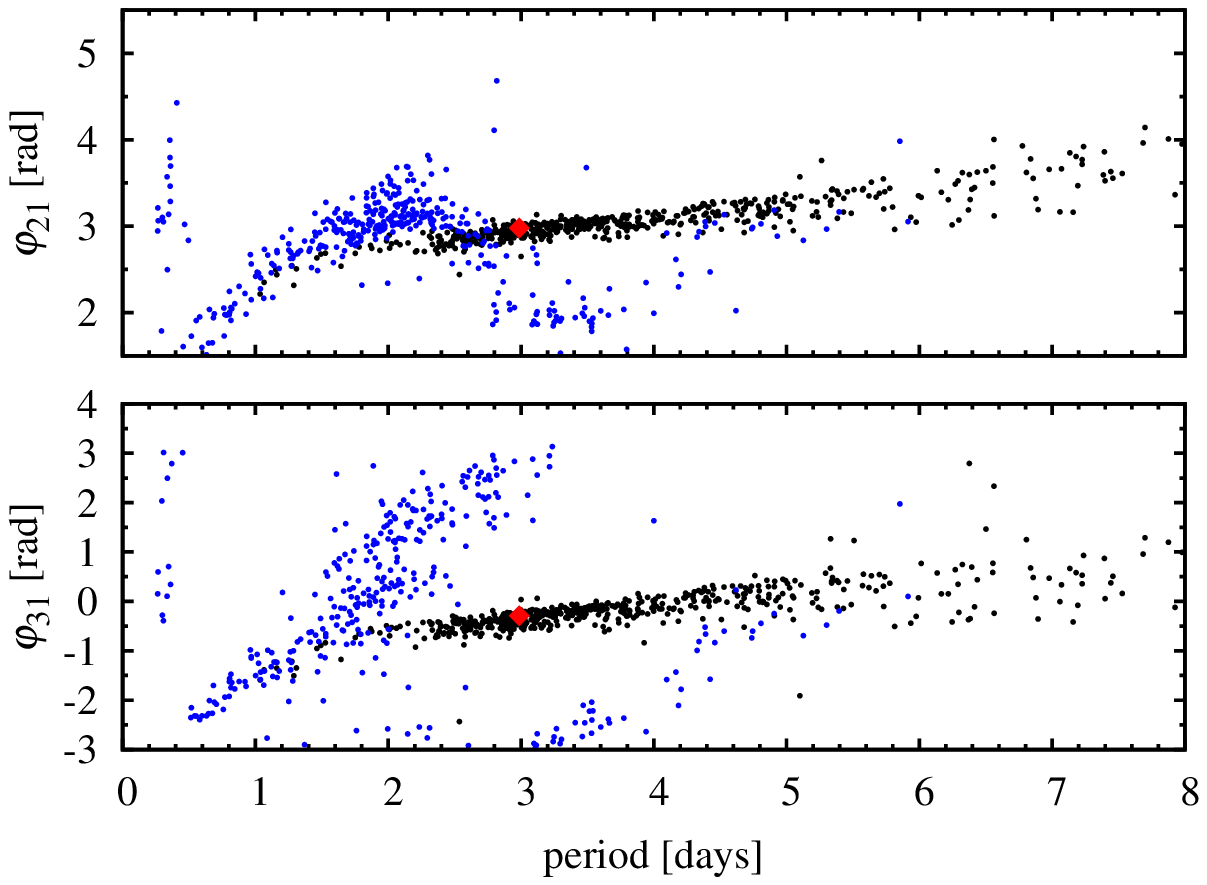}} \\
\caption{Fourier decomposition amplitudes and phases of the I-band light curve
of the Cepheid in the OGLE LMC562.05.9009 system. The symbols on the
Y-axes of the different panels have their usual meaning, and the data for the LMC Cepheids (small dots)
come from the OGLE Survey (Soszy{\'n}ski et al. 2008). The light of
the stable companion was subtracted from the Cepheid light curve before
the analysis. The location of the Cepheid on all panels clearly supports
fundamental mode pulsation, in agreement with the radius of the Cepheid
derived from our analysis (see text).
\label{fig:fourier}}
\end{center}
\end{figure}

\begin{figure}
\begin{center}
  \resizebox{\linewidth}{!}{\includegraphics{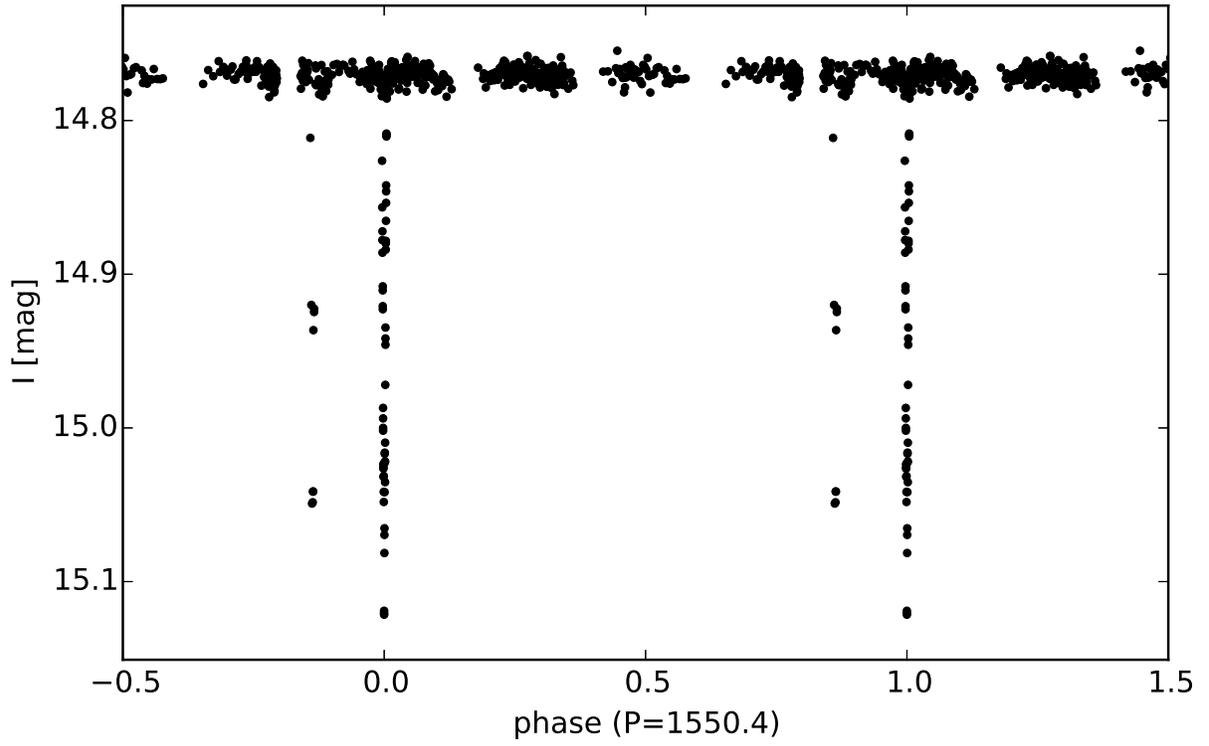}} \\
\caption{The I-band orbital light curve of the system determined from OGLE data, freed from the pulsations of the Cepheid. It is seen that both eclipses in this highly eccentric system are covered by the data.
\label{fig:lcec}}
\end{center}
\end{figure}


\newcommand{\mc}{\multicolumn}

\begin{deluxetable}{lcccc}
\tablecaption{Orbital solution for LMC562.05.9009}
\tablewidth{0pt}
\tablehead{
\colhead{Parameter} & \colhead{Solution 1} & \colhead{Solution 2} & \colhead{Solution 3} & \colhead{Unit}
}
\startdata
$T_0$           & 4230.67$^a$ & 4229.37$^a$ & 4233.50$^a$ &  days   \\
$\gamma$        &  296.12(9) &  296.21(10) &  296.34(10) &  km/s   \\ 
$K_1$           &   22.19(7)  &   22.24(7)  &   22.17(8)  &  km/s \\ 
$K_2$           &   22.78(7)  &   22.87(8)  &   22.98(8)  &  km/s \\ 
$e$             & 0.6113$^a$  & 0.6116$^a$  & 0.6150(15)  &  - \\
$\omega$        &  4.5$^a$    &    5.08(5)  &    3.90(5)  &  degrees  \\
$a \sin i$      &    1091(3)  &    1094(3)  &    1091(3)  &  $R_\odot$ \\ 
$m_1 \sin^3 i$  &    3.67(3)  &    3.70(3)  &    3.69(3)  &  $M_\odot$ \\ 
$m_2 \sin^3 i$  &    3.57(3)  &    3.60(3)  &    3.56(3)  &  $M_\odot$ \\ 
$q=m_2/m_1$     &    0.974(4) &    0.973(5) &    0.965(5) &  - \\ 
rms$_1$         &    0.35     &    0.36     &    0.35     &  km/s \\
rms$_2$         &    0.41     &    0.38     &    0.40     &  km/s 
\enddata
\label{tab:spec}
\vspace{-0.8cm}
\tablecomments{$T_0$ ($HJD - 2450000$ days) calculated from the epoch of the primary minimum $T_I = 3959.23 d$. Errors in the last significant digits are shown in parenthesis.
The last two rows show the rms scatter of the orbital radial velocities about the fitted curves, for the Cepheid ($1$) and its companion star ($2$).}
\tablenotetext{a}{ fixed value taken from the photometric solution}

\end{deluxetable}


\begin{figure}
\begin{center}
  \resizebox{\linewidth}{!}{\includegraphics{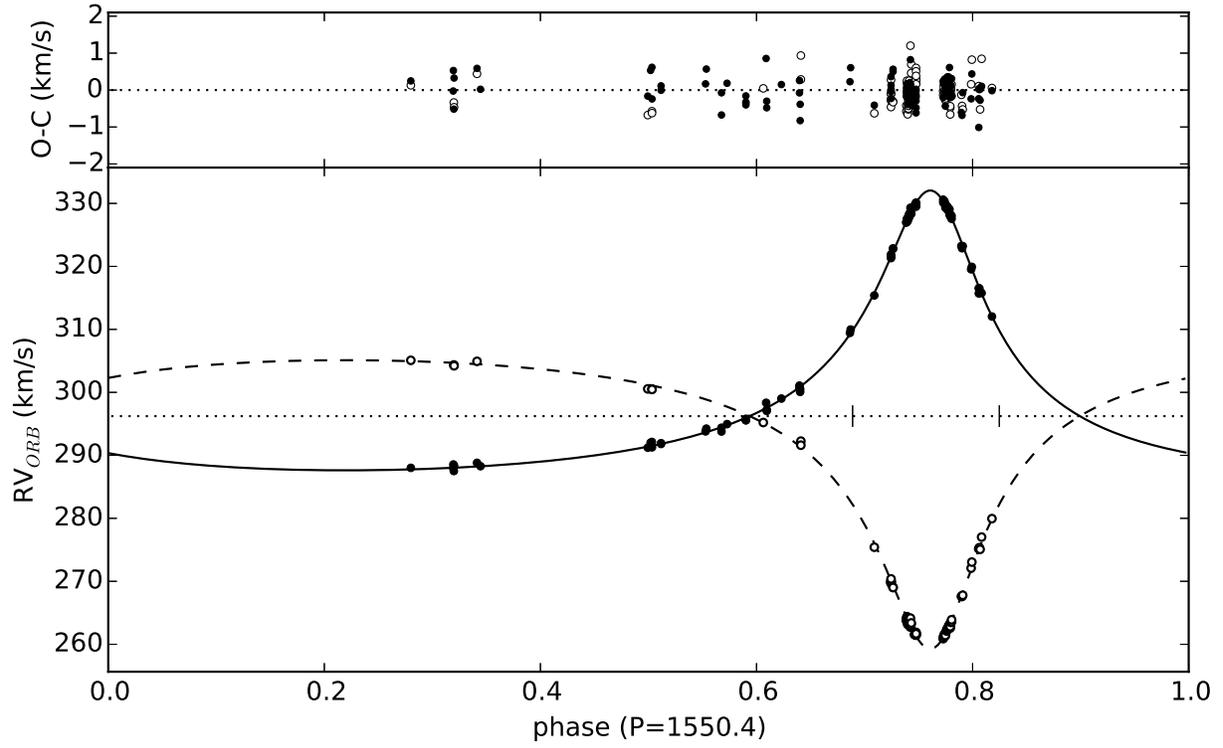}} \\
\caption{Orbital solution for LMC562.05.9009. Measured radial velocities of the Cepheid with the pulsations removed (filled circles) and of its non-pulsating companion 
(open circles) are shown. Small vertical lines mark the positions of the eclipses for this configuration.
\label{fig:rvorb}}
\end{center}
\end{figure}


\begin{figure}
\begin{center}
  \resizebox{\linewidth}{!}{\includegraphics{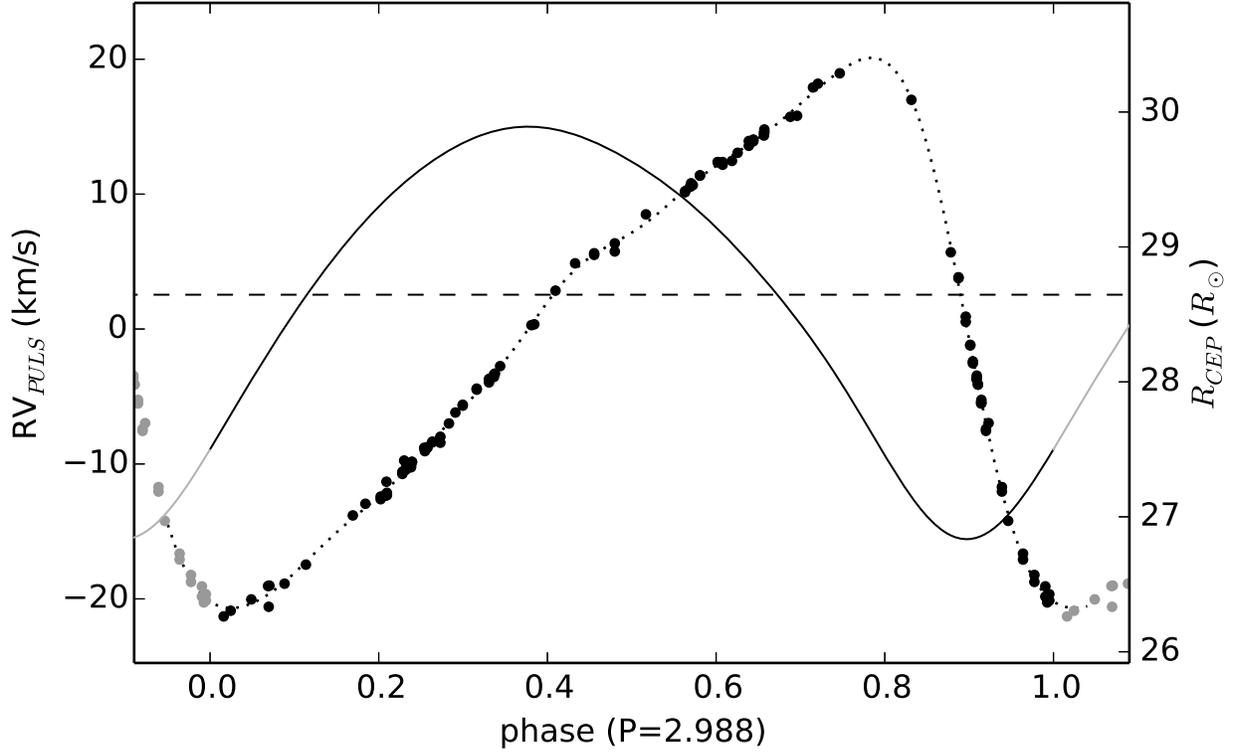}} \\
\caption{Pulsational radial velocity curve (black circles) and radius variation of the Cepheid over one pulsation cycle (solid line). To obtain the
pulsational radial velocity curve of the Cepheid,
its orbital motion was removed from the measured radial velocities. The full amplitude of the radius change is 3.04 R$_\odot$. The mean radius is marked
by the dashed line.
The grey datapoints are repeated to facilitate a better appreciation of the radial velocity curve (same for the radius variation curve).
\label{fig:rvpuls}}
\end{center}
\end{figure}


\begin{deluxetable}{lcccc}
\tablecaption{Photometric parameters of LMC562.05.9009 from the Monte Carlo simulations}
\tablewidth{0pt}
\tablehead{
\colhead{Parameter} & \colhead{Mean$_(S1)$} & \colhead{Solution 1} & \colhead{Mean$_(S2)$} & \colhead{Solution 2}
}
\startdata
$P_{orb} (d) $  &  -        &  1550.354(9)     &  -        &  1550.355(9)     \\
$T_{I}$ (d)     &  -        &  3959.227(17)    &  -        &  3959.225(17)    \\
$r_1$           &  0.02619  &  0.02514(18)$^a$ &  0.02599  &  0.02494(18)$^a$ \\
$r_2$           &  -        &  0.02427(16)     &  -        &  0.02422(17)     \\
$j_{21}(V)$     &  1.001    &  0.623(7)$^a$    &  0.991    &  0.617(8)$^a$    \\
$j_{21}(R_C)$   &  0.975    &  0.673(13)$^a$   &  0.957    &  0.661(13)$^a$   \\ 
$j_{21}(I_C)$   &  1.000    &  0.755(7)$^a$    &  0.984    &  0.742(8)$^a$    \\
$l_{3}(R_C)$    &  0.045    &  0.039(8)$^a$    &  0.045    &  0.039(8)$^a$    \\
$i$ ($^\circ$)  &  -        &  87.98(1)        &  -        &  87.99(1)        \\
$e$             &  -        &  0.61160(5)      &  -        &  0.6113(2)       \\
$\omega$ ($^\circ$)& -      &  5.08 (fixed)    &  -        &  4.5(4)          \\
$p$-factor      &  -        &  1.37(7)         &  -        &  1.37(7)         \\
\multicolumn{4}{l}{Derived quantities:}\\
$L_{21}(V)$     &  0.857    &  0.581(13)$^a$   &  0.858    &  0.582(13)$^a$   \\ 
$L_{21}(R_C)$   &  0.834    &  0.627(19)$^a$   &  0.830    &  0.624(19)$^a$   \\ 
$L_{21}(I_C)$   &  0.858    &  0.704(16)$^a$   &  0.854    &  0.700(16)$^a$   \\ 
\multicolumn{4}{l}{Additional information:}\\
rms ($V$)       &           & 0.0058           &           & 0.0058 \\ 
rms ($R_C$)     &           & 0.0052           &           & 0.0052 \\
rms ($I_C$)     &           & 0.0075           &           & 0.0075
\label{tab:photpar}
\enddata
\vspace{-0.8cm}
\tablecomments{Epoch of the primary eclipse $T_{I}$ is $HJD - 2450000$~d, $L_{21}$ is the light ratio of the components in every photometric band. Mean values
for each solution are given for parameters which change during the pulsation cycle.}
\tablenotetext{a}{ values correspond to a pulsation phase 0.0}
\end{deluxetable}
\begin{figure*}
\begin{center}
  \resizebox{0.32\linewidth}{!}{\includegraphics{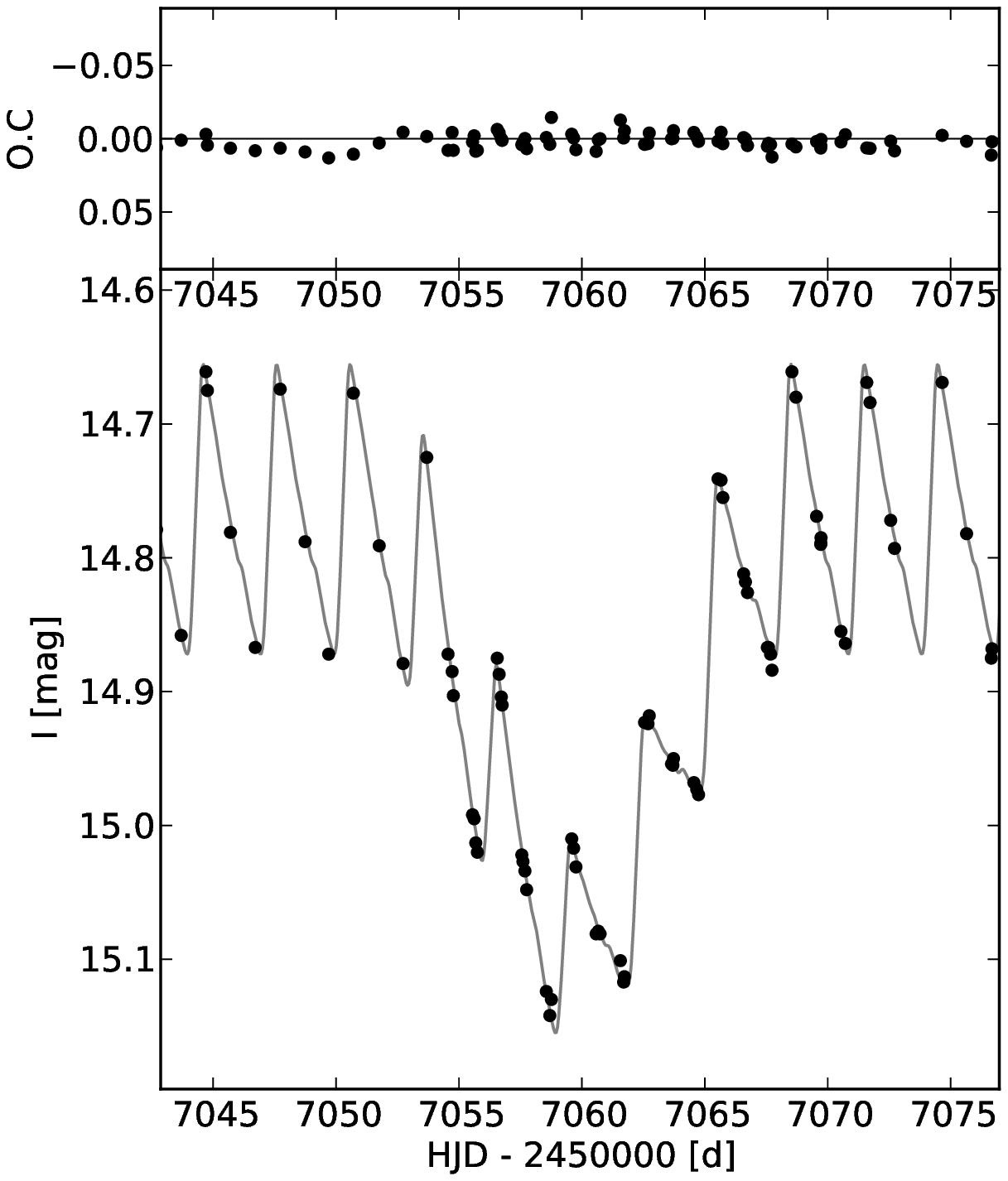}}
  \resizebox{0.32\linewidth}{!}{\includegraphics{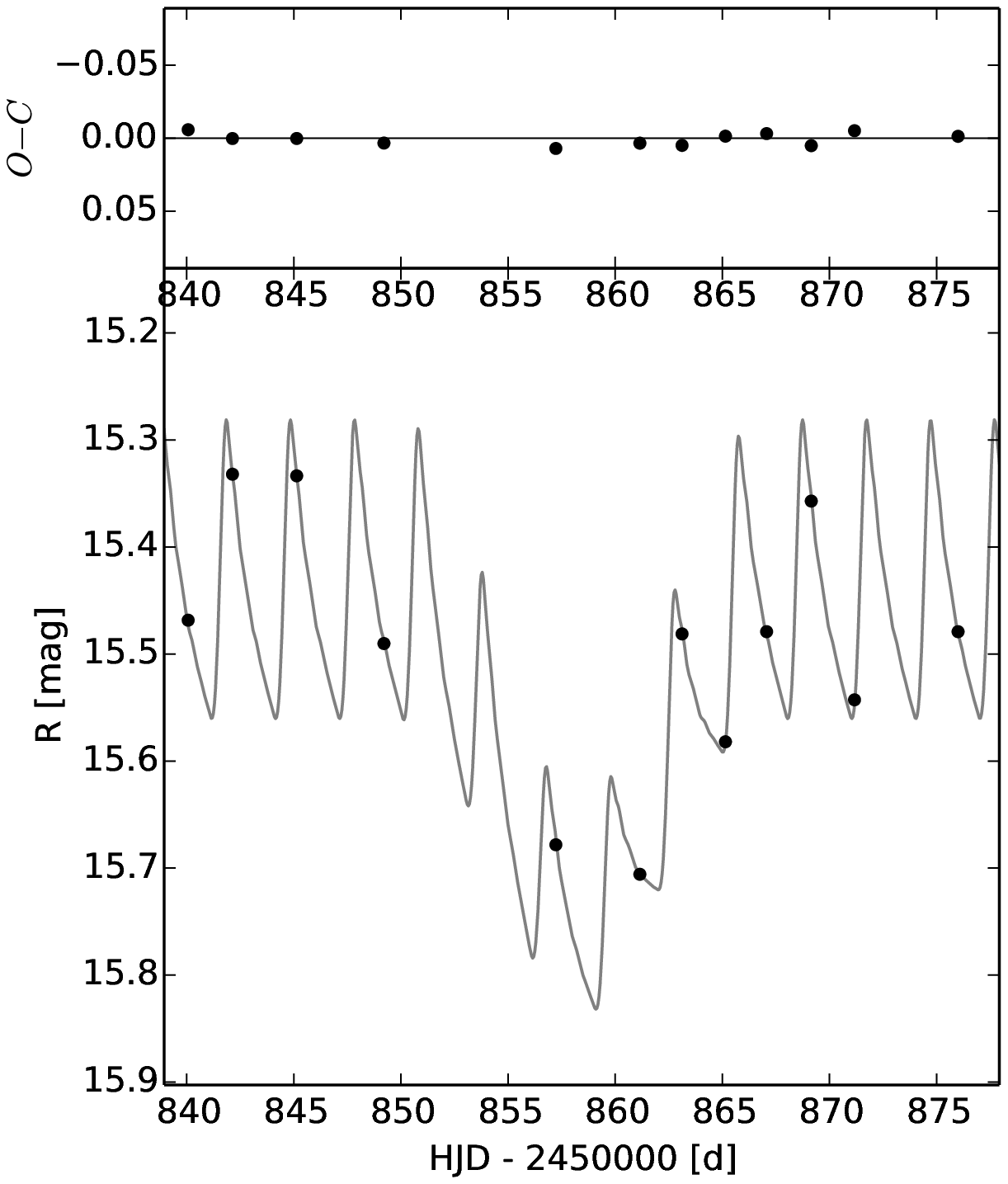}}
  \resizebox{0.32\linewidth}{!}{\includegraphics{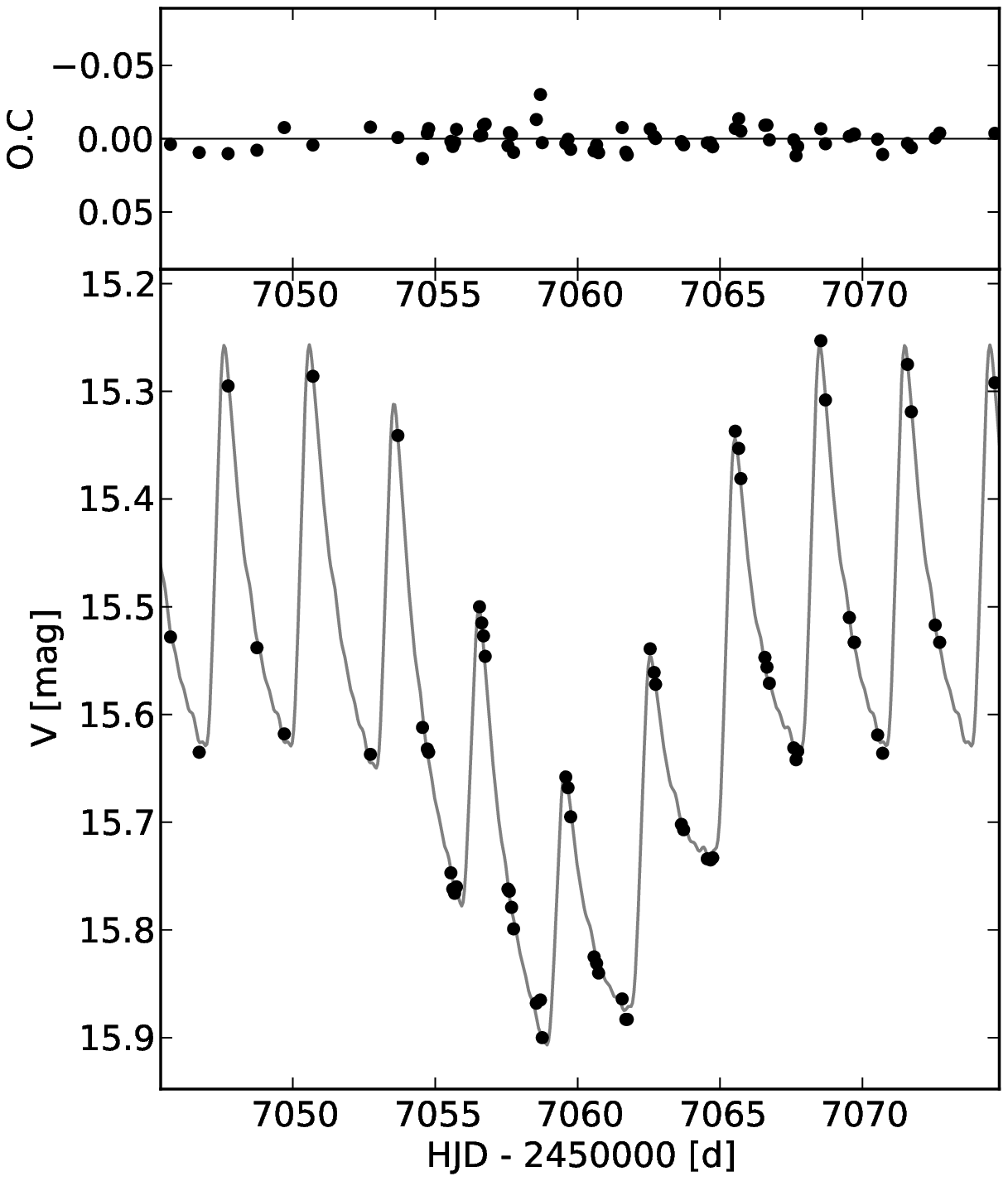}}
\caption{Close-up on the selected primary eclipses. The eclipse is caused by the transit of the companion star over the disc of the Cepheid. As expected, the pulsation amplitude 
is lower during the eclipses, due to the smaller contribution of the Cepheid to the total light. The {\em rms} scatter is about 0.008, 0.005 and 0.006 for $I_C$, $R_C$ and $V$, respectively.
\label{fig:ieclmodel}}
\end{center}
\end{figure*}


\begin{figure}
\begin{center}
  \resizebox{\linewidth}{!}{\includegraphics{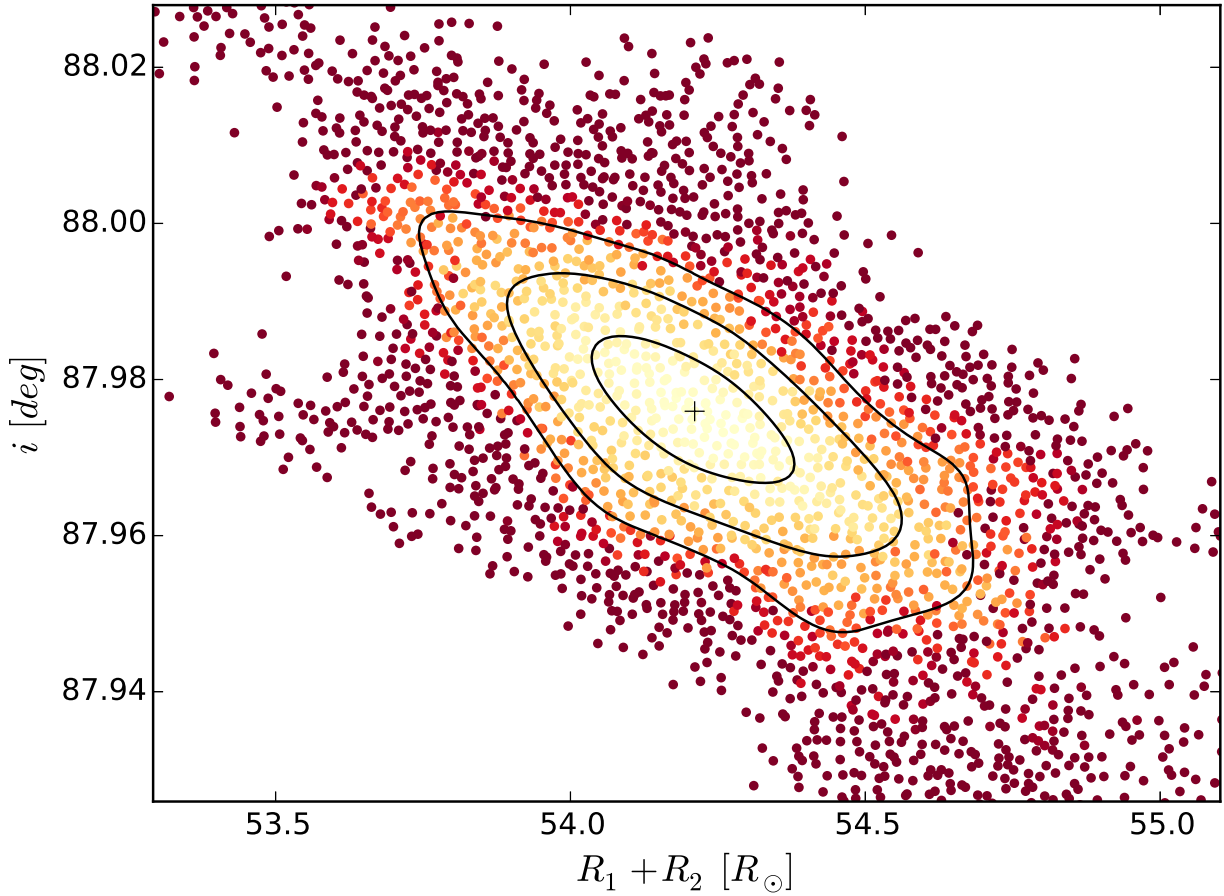}}
\caption{Correlation between the inclination and the sum of the star radii. The $\chi^2$ values are coded with color (higher values are darker). Solid lines represent 1-, 2- and 3-$\sigma$ levels for 
the two-parameter error estimation. The best model is marked with a cross.
\label{fig:corr_i}}
\end{center}
\end{figure}


\begin{figure}
\begin{center}
  \resizebox{\linewidth}{!}{\includegraphics{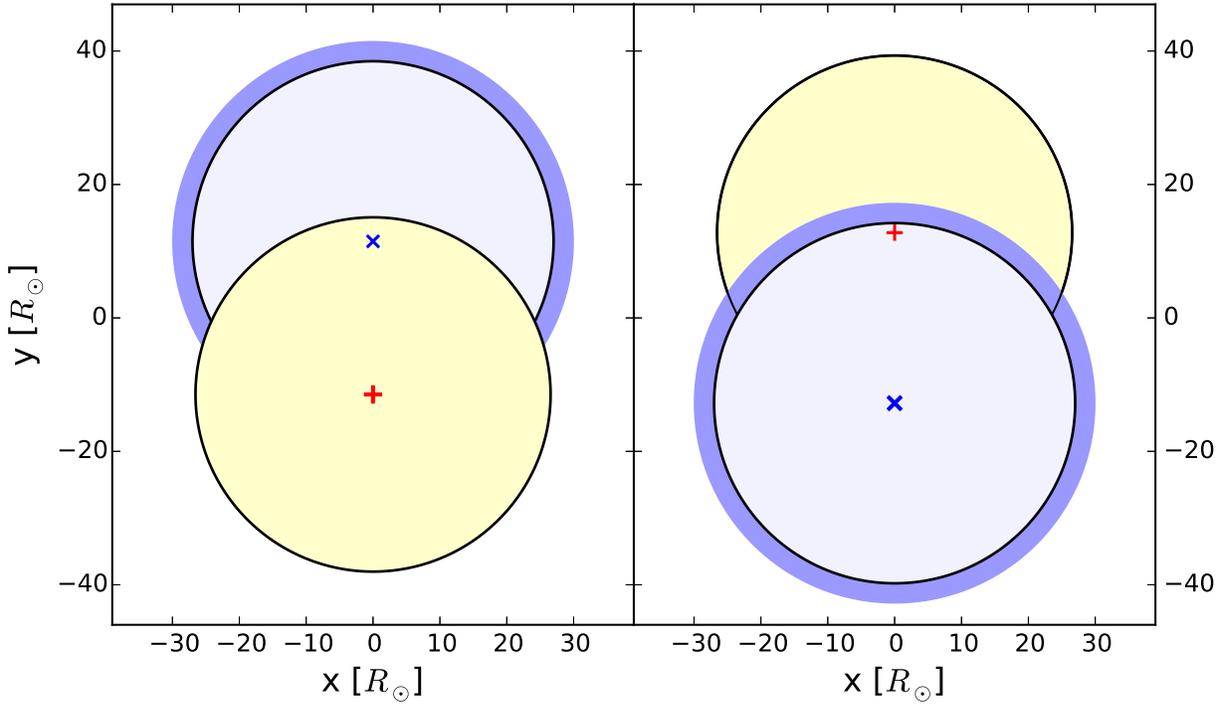}} \\
\caption{System configuration at the primary ({\em left}) and the secondary eclipse ({\em right}). The~Cepheid center is marked by a blue $\times$, its minimum radius by a 
solid line and the range of radius change by a blue transparent color. The companion center is marked by a red $+$ and the radius by the solid line. In both cases the width 
of the solid line represents the $\pm 1\sigma$ radius error. The distances between the stars are about 650 $R_\odot$ and 725 $R_\odot$ at the primary and secondary eclipse, respectively.
\label{fig:config}}
\end{center}
\end{figure}

\begin{figure}
\begin{center}
  \resizebox{\linewidth}{!}{\includegraphics[angle=-90]{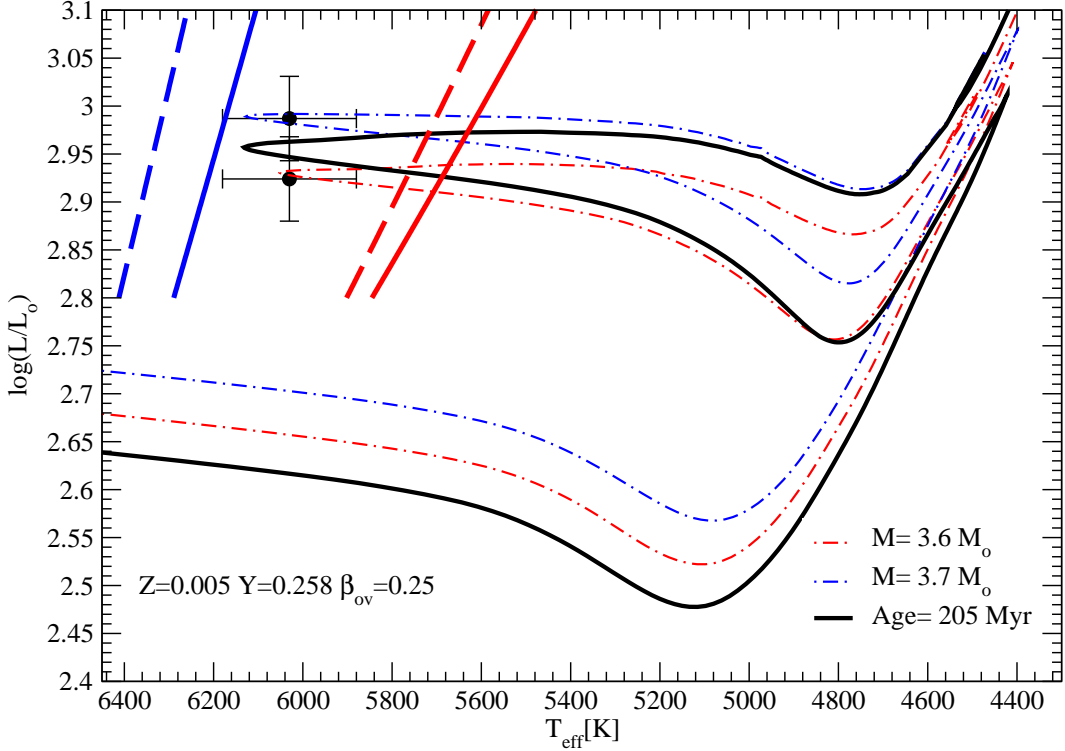}} \\
\caption{The observed locations of the classical Cepheid (upper black circle) and its stable red giant companion (lower black circle) on the luminosity-
effective temperature diagram. Blue and red lines show the blue and red
edges of the Cepheid instability strip, for metallicities of Z=0.008
(solid lines), and Z=0.004 (dashed lines). Both stars are, within the
observational uncertainties, located inside the theoretical instability
strip (Bono et al. 2005), for both metallicities. The blue and red
pointed lines show the evolutionary tracks for the Cepheid, and its
slightly less massive companion, respectively. The solid black line
is the isochrone computed for an age of 205 Myr which fits the positions
of both stars, within the uncertainties (for more details on the
calculations, see text).
\label{fig:9009_tracks}}
\end{center}
\end{figure}

\begin{deluxetable}{lccc}
\tablecaption{Physical properties of the component stars in LMC562.05.9009}
\tablewidth{0pt}
\tablehead{
 \colhead{Parameter} & \colhead{Primary (Cepheid)} & \colhead{Secondary} & \colhead{Unit}}
\startdata
pulsation period & 2.987846(1) &    -   & days    \\
mass             & 3.70(3)     & 3.60(3)  & $M_\odot$ \\ 
radius           & 28.6(2)     & 26.6(2)  & $R_\odot$ \\ 
$\log g$         & 2.129(6)    & 2.146(6) & cgs       \\
temperature & 6030(150) & 6030(150) & K \\
$\log{L}$               & 2.987(44)& 2.924(44)& $L_\odot$ \\ 
$V$              & 15.837      & 15.999   & mag     \\
$(V\!-\!I_C)$        &  0.585      &  0.584  & mag \\
$(V\!-\!K)$           & 1.316       & 1.327 & mag \\
$E(B\!-\!V)$    & \multicolumn{2}{c}{0.106(27)} & mag\\    
\label{tab:abs}
\enddata
\vspace{-0.8cm}
\tablecomments{ For the Cepheid all variable quantities are the mean values over the pulsation cycle. Magnitudes and colors
are corrected for the extinction.}
\end{deluxetable}

\begin{deluxetable}{@{}lccccccc@{}}
\tabletypesize{\small}
\tablecaption{Magnitudes of the LMC562.05.9009 system} 
\label{tab:magnitud}
\tablewidth{0pt}
\tablehead{
\colhead{Light} &\colhead{Expected}& \colhead{Observed} & \multicolumn{2}{c}{Extinction}& \colhead{Dereddened} & \colhead{Absolute$^{a}$}& \colhead{Bolometric$^{b}$}\\
\colhead{Source} & \colhead{mag.}& \colhead{mag.}& \colhead{differ.}& \colhead{total}& \colhead{mag.} & \colhead{mag.} & \colhead{mag.}}
\startdata  
  \multicolumn{8}{c}{$V$-band}  \\
total      		& 15.54	& 15.49	&$-0.05$	&0.33$^{g}$&15.16	&-		& -  \\ 
Cepheid            & 16.22$^{c}$	& 16.16	&-		&-		&15.84	&-2.66	&-2.71 \\
companion      	&16.38$^{e}$& 16.33&-		&-		&16.00	&-2.49	&-2.54 \\ 
\hline
\multicolumn{8}{c}{$I_C$-band}  \\
total      		& 14.80	& 14.77	&$-0.03$	&0.19$^{g}$&14.58	&-		& - \\ 
Cepheid             & 15.47$^{c}$	& 15.45	&-		&-		&15.25	&-3.24	& - \\
companion      	&15.64$^{e}$& 15.61&-		&-		&15.42	&-3.08	& - \\ 
\hline
\multicolumn{8}{c}{$K$-band}  \\
total      		& -	& 13.88	&-		&0.04$^{h}$&13.84	&-		& - \\ 
Cepheid            & - 	& 14.55	&-		&-		&14.52$^{d}$	&-3.97	& - \\
companion      	& -	& 14.71	&-		&-		&14.67	&-3.82	& - \\ 
\enddata
\newline
\tablecomments{Flux-weighted means are given for the system (total) and the Cepheid.}
\tablenotetext{a}{assuming a distance modulus to the LMC of 18.49 (Pietrzy{\'n}ski et al. 2013)}
\tablenotetext{b}{bolometric corrections from Worthey\&Lee (2011)}
\tablenotetext{c}{from the observed (reddened) relation for FU classical Cepheids of Soszy{\'n}ski et al.(2008)}
\tablenotetext{d}{from the extinction-corrected relation for FU classical Cepheids of Ripepi et al.(2012)}
\tablenotetext{e}{from light ratio of the components - Solution 1 in Table~\ref{tab:photpar}}
\tablenotetext{g}{added foreground and mean internal reddening of the LMC: $A_V=0.38$ mag and $A_I=0.22$ mag}
\tablenotetext{h}{from the relation $A_K=0.34\cdot E(B-V)$}
\end{deluxetable}

\end{document}